\documentclass[11pt]{article}
\usepackage{epsfig}

\def\beq{\begin{equation}}
\def\eeq#1{\label{#1}\end{equation}}
\def\eeqn{\end{equation}}

\def\beqa{\begin{eqnarray}}
\def\eeqa#1{\label{#1}\end{eqnarray}}
\def\eeqan{\end{eqnarray}}

\let\bar=\overbar

\def\D{{\cal D}}

\def\Dslash{\not{\hbox{\kern-4pt $D$}}}
\def\dslash{\not{\hbox{\kern-2pt $\del$}}}

\def\msb{{\bar{\ssstyle M \kern -1pt S}}}

\def\Title#1{\begin{center} {\Large {\bf #1} } \end{center}}
\def\barD{\overline D{}^0}

\def\DDbar{D^0-\overline D{}^0}

\def\D0bar{\overline D{}^0}
\def\K0bar{\overline K{}^0}

\def\3bar{\overline{3}}
\def\sixbar{\overline{6}}

\def\15bar{\overline{15}}
\def\24bar{\overline{24}}
\def\42bar{\overline{42}}
\def\60bar{\overline{60}}
\def\cO{{\cal O}}

\def\beq{\begin{equation}}
\def\eeq{\end{equation}}
\def\beqa{\begin{eqnarray}}
\def\eeqa{\end{eqnarray}}

\begin{document}
\begin{flushright}
WSU-HEP-0207\\hep-ph/0207212
\end{flushright}

\Title{$D^0$-$\bar{D^0}$ mixing\footnote{Invited talk given at {\it 
Flavor Physics and CP Violation (FPCP)}, Philadelphia, PA, May 2002 }}

\bigskip\bigskip


\begin{raggedright}  

{\it Alexey A. Petrov\index{Petrov, A.A.}\\
Department of Physics and Astronomy\\
Wayne State University\\
Detroit, MI 48201}
\bigskip\bigskip
\end{raggedright}

\section{Introduction}

One of the most important motivations for studies of $\DDbar$ mixing 
is the possibility of observing a signal from new physics which can 
be separated from the one generated by the Standard Model (SM) interactions.
The $\DDbar$ mixing proceeds extremely slowly, which in the 
Standard Model is usually attributed to the absence of superheavy 
quarks destroying GIM cancelations~\cite{Petrov:1997ch}.
The low energy effect of new physics particles can be naturally
written in terms of a series of local operators of increasing
dimension generating $\Delta C = 2$ transitions. These operators,
as well as the one loop Standard Model effects, generate contributions 
to the effective operators that change $D^0$ state into $\barD$ state
leading to the mass eigenstates
\begin{equation} \label{definition1}
| D_{^1_2} \rangle =
p | D^0 \rangle \pm q | \bar D^0 \rangle,
\end{equation}
where the complex parameters $p$ and $q$ are obtained from diagonalizing 
the $D^0-\barD$ mass matrix. The mass and width splittings between these 
eigenstates are parameterized by
\begin{eqnarray} \label{definition}
x \equiv (m_2-m_1)/{\Gamma}, ~~
y \equiv (\Gamma_2 - \Gamma_1)/({2 \Gamma}),
\end{eqnarray}
where $m_{1,2}$ and $\Gamma_{1,2}$ are the masses and widths of
$D_{1,2}$  and the mean width and mass are $\Gamma=(\Gamma_1+\Gamma_2)/2$ 
and $m=(m_1+m_2)/2$. Since $y$ is constructed from the decays of $D$ into 
physical states, it should be dominated by the Standard Model contributions, 
unless new physics significantly modifies $\Delta C=1$ interactions.

Presently, experimental information about the $\DDbar$ mixing parameters 
$x$ and $y$ comes from the time-dependent analyses that can roughly be divided
into two categories. First, more traditional studies look at the time
dependence of $D \to f$ decays, where $f$ is the final state that can be
used to tag the flavor of the decayed meson. The most popular is the
non-leptonic doubly Cabibbo suppressed decay (DCSD) $D^0 \to K^+ \pi^-$.
Time-dependent studies allow one to separate the DCSD from the mixing 
contribution $D^0 \to \D0bar \to K^+ \pi^-$,
\begin{eqnarray}\label{Kpi}
\Gamma[D^0(t) \to K^+ \pi^-]
=e^{-\Gamma t}|A_{K^-\pi^+}|^2 \qquad\qquad\qquad\qquad\qquad\qquad\qquad\qquad\qquad
\nonumber \\
\times ~\left[
R+\sqrt{R}R_m(y'\cos\phi-x'\sin\phi)\Gamma t
+\frac{R_m^2}{4}(y^2+x^2)(\Gamma t)^2
\right],
\end{eqnarray}
where $R$ is the ratio of DCS and Cabibbo favored (CF) decay rates and 
$q/p=R_m e^{i \phi}$. Since $x$ and $y$ are small, the best constraint 
comes from the linear terms in $t$ that are also {\it linear} in $x$ and $y$.
A direct extraction of $x$ and $y$ from Eq.~(\ref{Kpi}) is not possible due 
to unknown relative strong phase $\delta$ of DCS and CF amplitudes~\cite{Falk:1999ts}, 
as $x'=x\cos\delta+y\sin\delta$, $y'=y\cos\delta-x\sin\delta$. This phase can be 
measured independently~\cite{GGR}. The corresponding formula can also be written
for $\barD$ decay with $x' \to -x'$ and $R_m \to R_m^{-1}$~\cite{Bergmann:2000id}.

Second, $D^0$ mixing can be measured by comparing the lifetimes 
extracted from the analysis of $D$ decays into the CP-even and CP-odd final states. 
This study is also sensitive to a {\it linear} function of $y$ via
\beq
\frac{\tau(D \to K^-\pi^+)}{\tau(D \to K^+K^-)}-1=
y \cos \phi - x \sin \phi \left[\frac{R_m^2-1}{2}\right].
\eeq
Time-integrated studies of the semileptonic transitions are sensitive
to the {\it quadratic} form $x^2+y^2$ and at the moment are not competitive with the
analyses discussed above. The construction of a new tau-charm factory at Cornell
will introduce other time-independent methods that are sensitive to a 
linear function of $y$~\cite{AtwoodPetrov}. 

The current experimental upper bounds on $x$ and $y$ are on the order of 
a few times $10^{-2}$, and are expected to improve significantly in the coming
years.  To regard a future discovery of nonzero $x$ or $y$ as a signal for new 
physics, we would need high confidence that the Standard Model predictions lie
well below the present limits.  As was recently shown~\cite{Falk:2001hx}, 
in the Standard Model $x$ and $y$ are generated only at second order in $SU(3)$ 
breaking, 
\beq
x\,,\, y \sim \sin^2\theta_C \times [SU(3) \mbox{ breaking}]^2\,,
\eeq
where $\theta_C$ is the Cabibbo angle.  Therefore, predicting the
Standard Model values of $x$ and $y$ depends crucially on estimating the 
size of $SU(3)$ breaking.  Although $y$ is expected to be determined
by Standard Model processes, its value nevertheless affects significantly 
the sensitivity to new physics of experimental analyses of $D$ 
mixing~\cite{Bergmann:2000id}.

\section{Theoretical Expectations}

Theoretical predictions of $x$ and $y$ within and beyond
the Standard Model span several orders of magnitude (see Ref.~\cite{Nelson:1999fg}).
Roughly, there are two approaches, neither of which give very reliable
results because $m_c$ is in some sense intermediate between heavy and
light.  The ``inclusive'' approach is based on the operator
product expansion (OPE).  In the $m_c \gg \Lambda$ limit, where
$\Lambda$ is a scale characteristic of the strong interactions, $\Delta
M$ and $\Delta\Gamma$ can be expanded in terms of matrix elements of local
operators~\cite{Inclusive}.  Such calculations yield $x,y < 10^{-3}$.  
The use of the OPE relies on local quark-hadron duality, 
and on $\Lambda/m_c$ being small enough to allow a truncation of the series
after the first few terms.  The charm mass may not be large enough for these 
to be good approximations, especially for nonleptonic $D$ decays.
An observation of $y$ of order $10^{-2}$ could be ascribed to a
breakdown of the OPE or of duality,  but such a large
value of $y$ is certainly not a generic prediction of OPE analyses.
The ``exclusive'' approach sums over intermediate hadronic
states, which may be modeled or fit to experimental data~\cite{Exclusive}.
Since there are cancellations between states within a given $SU(3)$
multiplet, one needs to know the contribution of each state with high 
precision. However, the $D$ is not light enough that its decays are dominated
by a few final states.  In the absence of sufficiently precise data on many decay 
rates and on strong phases, one is forced to use some assumptions.  While most 
studies find $x,y < 10^{-3}$, Refs.~\cite{Exclusive} obtain $x$ and 
$y$ at the $10^{-2}$ level by arguing that $SU(3)$ violation is of order
unity, but the source of the large $SU(3)$ breaking is not made explicit.

In what follows we first prove that $D^0-\D0bar$ mixing arises only at 
{\it second} order in $SU(3)$ breaking effects.  The proof is valid when
$SU(3)$ violation enters perturbatively. This would not be so, for
example, if $D$ transitions were dominated by a single narrow 
resonance close to threshold~\cite{Falk:2001hx,Golowich:1998pz}. Then we argue that 
reorganization of ``exclusive'' calculation by explicitly building 
$SU(3)$ cancellations into the analysis naturally leads to values of 
$y \sim 1\%$ if only one source of $SU(3)$ breaking (phase space) is taken 
into account.

The quantities $M_{12}$ and $\Gamma_{12}$ which determine $x$ and $y$
depend on matrix elements 
$\langle\D0bar|\, {\cal H}_w {\cal H}_w\, |D^0\rangle\,$,
where ${\cal H}_w$ denote the $\Delta C=-1$ part of the weak Hamiltonian.  
Let $D$ be the field operator that creates a $D^0$ meson and annihilates a
$\D0bar$.  Then the matrix element, whose $SU(3)$ flavor group theory properties
we will study, may be written as
\beq\label{melm}
      \langle 0|\, D\, {\cal H}_w {\cal H}_w \,D\, |0 \rangle\,.
\eeq
Since the operator $D$ is of the form $\bar cu$, it transforms in the
fundamental representation of $SU(3)$, which we will represent with a
lower index, $D_i$.  We use a convention in which the correspondence between 
matrix indices and quark flavors is $(1,2,3)=(u,d,s)$.  The only nonzero 
element of $D_i$ is $D_1=1$.  The $\Delta C=-1$
part of the weak Hamiltonian has the flavor structure $(\bar q_ic)(\bar
q_jq_k)$, so its matrix representation is written with a fundamental
index and two antifundamentals, $H^{ij}_k$.  This operator is a sum of irreps
contained in the product $3 \times \3bar \times \3bar =
\15bar + 6 + \3bar + \3bar$.  In the limit in which the third generation is
neglected, $H^{ij}_k$ is traceless, so only the $\15bar$ 
and 6 representations appear.  That is, the
$\Delta C=-1$ part of ${\cal H}_w$ may be decomposed as ${1\over2} 
(\cO_{\15bar} + \cO_6)$,
where
\beqa
\cO_{\15bar} &=& (\bar sc)(\bar ud) + (\bar uc)(\bar sd)
    + s_1(\bar dc)(\bar ud) + s_1(\bar uc)(\bar dd)\nonumber\\
&&{} - s_1(\bar sc)(\bar us) - s_1(\bar uc)(\bar ss)
    - s_1^2(\bar dc)(\bar us) - s_1^2(\bar uc)(\bar ds) \,, \nonumber\\
\cO_6 &=& (\bar sc)(\bar ud) - (\bar uc)(\bar sd)
    + s_1(\bar dc)(\bar ud) - s_1(\bar uc)(\bar dd)\nonumber\\
&&{} - s_1(\bar sc)(\bar us) + s_1(\bar uc)(\bar ss)
    - s_1^2(\bar dc)(\bar us) + s_1^2(\bar uc)(\bar ds) \,,
\eeqa
and $s_1=\sin\theta_C$.  The matrix representations
$H(\15bar)^{ij}_k$ and $H(6)^{ij}_k$ have nonzero elements
\begin{equation}
\begin{tabular}{rll}
$H(\15bar)^{ij}_k:\qquad$
    &  $H^{13}_2 = H^{31}_2=1$\,,  &  $H^{12}_2 = H^{21}_2 = s_1$\,,\\
&  $H^{13}_3 = H^{31}_3 = -s_1$\,,  &  $H^{12}_3 =
H^{21}_3=-s_1^2$\,,\\[4pt]
$H(6)^{ij}_k:\qquad$
    &  $H^{13}_2 = -H^{31}_2=1$\,,  &  $H^{12}_2 = -H^{21}_2 = s_1$\,,\\
&  $H^{13}_3 = -H^{31}_3 = -s_1$\,,$\qquad$
    &  $H^{12}_3 = -H^{21}_3 = -s_1^2$\,.
\end{tabular}
\end{equation}
We introduce $SU(3)$ breaking through the quark mass operator ${\cal
M}$, whose matrix representation is $M^i_j={\rm diag}(m_u,m_d,m_s)$
as being in the adjoint representation to induce $SU(3)$ violating effects.  
We set $m_u=m_d=0$ and let $m_s\ne0$ be the only $SU(3)$ violating parameter.  
All nonzero matrix elements built out of $D_i$, $H^{ij}_k$ and $M^i_j$ must be 
$SU(3)$ singlets.

We now prove that $D^0-\D0bar$ mixing arises only at second order in
$SU(3)$ violation, by which we mean second order in $m_s$.  First, we note that
the pair of $D$ operators is symmetric, and so the product $D_iD_j$
transforms as a 6 under $SU(3)$.  Second, the pair of ${\cal H}_w$'s is also symmetric, 
and the product $H^{ij}_kH^{lm}_n$ is in one of the reps which
appears in the product
\beqa
\left[ (\15bar+6)\times(\15bar+6) \right]_S &=&
    (\15bar\times\15bar)_S +(\15bar\times 6)+(6\times 6)_S \\*
&=& (\60bar+\24bar+15+15'+\sixbar) + (42+24+15+\sixbar+3)
    + (15'+\sixbar)\,. \nonumber
\eeqa
A direct computation shows that only three of these
representations actually appear in the decomposition of ${\cal H}_w{\cal H}_w$.  They 
are the $\60bar$, the 42, and the $15'$ (actually twice, but with the same nonzero elements
both times).  So we have product operators of the form (the subscript denotes the 
representation of $SU(3)$)
\beqa
    DD = {\cal D}_6\,, ~~~~
    {\cal H}_w {\cal H}_w = \cO_{\60bar}+\cO_{42}+\cO_{15'}\,.
\eeqa
Since there is no $\sixbar$ in the decomposition of ${\cal H}_w{\cal 
H}_w$, there is no $SU(3)$ singlet which can be made with ${\cal D}_6$,  
and no $SU(3)$ invariant matrix element of the form (\ref{melm}) can be formed.  
This is the well known result that $D^0-\D0bar$ mixing is 
{\it prohibited by $SU(3)$ symmetry}.
Now consider a single insertion of the $SU(3)$ violating spurion ${\cal M}$.
The combination ${\cal D}_6{\cal M}$ transforms as $6\times
8=24+\15bar+6+\3bar$. There is still no invariant to be made
with ${\cal H}_w{\cal H}_w$, thus $D^0-\D0bar$ mixing is {\it not 
induced at first order in $SU(3)$ breaking}.
With two insertions of ${\cal M}$, it becomes possible to make an $SU(3)$
invariant.  The decomposition of ${\cal D}{\cal M}{\cal M}$ is
\beqa
6\times(8\times 8)_S = 6\times(27+8+1)
    = (60+\42bar+24+\15bar+\15bar'+6) + (24+\15bar+6+\3bar) + 6\,.
\eeqa
There are three elements of the $6\times 27$ part which can give invariants
with  ${\cal H}_w{\cal H}_w$.  Each invariant yields a contribution to 
$D^0-\D0bar$ mixing proportional to $s_1^2m_s^2$. Thus, $\DDbar$ mixing arises 
only at {\it second order} in the $SU(3)$ violating parameter $m_s$.

We now turn to the contributions to $y$ from on-shell final states, which result
from every common decay product of $D^0$ and $\D0bar$.  In the $SU(3)$ limit, 
these contributions cancel when one sums over complete $SU(3)$ multiplets in 
the final state.  The cancellations
depend on $SU(3)$ symmetry both in the decay matrix elements and in the
final state phase space.  While there are $SU(3)$ violating
corrections to both of these, it is difficult to compute the $SU(3)$
violation in the matrix elements in a model independent manner. Yet, with 
some mild assumptions about the momentum dependence of the matrix elements, the 
$SU(3)$ violation in the phase space depends only on the final particle masses and 
can be computed. We estimate the contributions to $y$ solely from 
$SU(3)$ violation in the phase space. We find that this source of $SU(3)$ 
violation can generate $y$ of the order of a few percent.

The mixing parameter $y$ may be written in terms of the matrix elements
for common final states for $D^0$ and $\D0bar$ decays,
\beq
y = {1\over\Gamma} \sum_n \int [{\rm P.S.}]_n\,
    \langle \D0bar|\,{\cal H}_w\,|n \rangle \langle n|\,{\cal H}_w\,|D^0
\rangle\,,
\eeq
where the sum is over distinct final states $n$ and the integral is over
the phase space for state $n$.  Let us now perform the phase space integrals
and restrict the sum to final states $F$ which  transform within a
single $SU(3)$ multiplet $R$.  The result is a  contribution to $y$ of
the form
\beq
    {1\over\Gamma}\, \langle\D0bar|\,{\cal H}_w
     \bigg\{ \eta_{CP}(F_R)\sum_{n\in  F_R}
    |n\rangle \rho_n\langle n| \bigg\} {\cal H}_w\,|D^0\rangle\,,
\eeq
where $\rho_n$ is the phase space available to the state $n$, 
$\eta_{CP}=\pm1$~\cite{Falk:2001hx}.  In the
$SU(3)$ limit, all the $\rho_n$ are the same for $n\in F_R$, and the quantity 
in braces above is an $SU(3)$ singlet.  Since the $\rho_n$ depend only on the 
known masses of the particles in the state $n$, incorporating the true values
of $\rho_n$ in the sum is a calculable source of $SU(3)$ breaking.

This method does not lead directly to a calculable contribution to $y$,
because
the matrix elements $\langle n|{\cal H}_w|D^0\rangle$ and
$\langle\D0bar|{\cal H}_w|n\rangle$
are not known.  However, $CP$ symmetry, which in the Standard Model and
almost all scenarios of new physics is to an excellent approximation
conserved in $D$ decays, relates $\langle\D0bar|{\cal H}_w|n\rangle$ to
$\langle
D^0|{\cal H}_w|\overline{n}\rangle$. Since $|n\rangle$ and
$|\overline{n}\rangle$ are
in a common $SU(3)$ multiplet,  they are determined by a single effective
Hamiltonian. Hence the ratio
\beq\label{yfr}
   y_{F,R} = {\sum_{n\in F_R} \langle\D0bar|\,{\cal H}_w|n\rangle \rho_n
    \langle n|{\cal H}_w\,|D^0\rangle \over
    \sum_{n\in F_R} \langle D^0|\,{\cal H}_w |n\rangle \rho_n
    \langle n|{\cal H}_w\,|D^0\rangle}
   = {\sum_{n\in F_R} \langle\D0bar|\,{\cal H}_w|n\rangle \rho_n
    \langle n|{\cal H}_w\,|D^0\rangle \over \sum_{n\in F_R}
   \Gamma(D^0\to n)}
\eeq
is calculable, and represents the value which $y$ would take if elements
of $F_R$ were the only channel open for $D^0$ decay.  To get a true
contribution
to $y$, one must scale $y_{F,R}$ to the total branching  ratio to all the
states in $F_R$.  This is not trivial, since a given physical final state
typically decomposes into a sum over more than one multiplet $F_R$.  The
numerator of $y_{F,R}$ is of order $s_1^2$ while the  denominator is of
order 1, so with large $SU(3)$ breaking in the phase space the natural size 
of $y_{F,R}$ is 5\%.
Indeed, there are other $SU(3)$ violating effects, such as in matrix elements 
and final state interaction phases.  Here we assume that
there is no cancellation with other sources of $SU(3)$ breaking, or
between the various multiplets which occur in $D$ decay, that would
reduce our result for $y$ by an order of magnitude.
This is equivalent to assuming that the $D$ meson is not
heavy enough for duality to enforce such cancellations. Performing the 
computations of $y_{F,R}$~\cite{Falk:2001hx},
we see that effects at the level of a few percent are quite generic.  
Our results are  summarized in Table~\ref{ytwobody}. Then, $y$ can be
formally constructed from the individual $y_{F,R}$ by
weighting them by their $D^0$ branching ratios,
\beq\label{ycombine}
    y = {1\over\Gamma} \sum_{F,R}\, y_{F,R}
    \bigg[\sum_{n\in F_R}\Gamma(D^0\to n)\bigg]\,.
\eeq
However, the data on $D$ decays are neither abundant nor precise enough
to disentangle the decays to the various $SU(3)$ multiplets, especially
for the three- and four-body final states.  Nor have we computed  $y_{F,R}$ for
all or even most of the available representations.   Instead, we can only
estimate individual contributions to $y$ by  assuming that the representations
for which we know $y_{F,R}$ to be  typical for final states with a given
multiplicity, and then to scale to  the total branching ratio to those final states.
The total branching  ratios of $D^0$ to two-, three- and four-body final
states can be extracted from Ref.~\cite{PDG}. Rounding to the nearest 5\% to emphasize 
the uncertainties in these numbers, we conclude that the branching fractions for 
$PP$, $(VV)_{\mbox{$s$-wave}}$, $(VV)_{\mbox{$d$-wave}}$ and $3P$ approximately 
amount to 5\%, while the branching ratios for $PV$ and $4P$ are of the order 
of 10\% \cite{Falk:2001hx}.

\begin{table}
\begin{center}
\begin{tabular}{|@{~~~}lc|c|c|} \hline
\multicolumn{2}{|c|}{~~Final state representation~~~}  &
     ~~~$y_{F,R}/s_1^2$~~~ & ~~~$y_{F,R}\ (\%)$~~~  \\ \hline\hline
    $PP$  &  $8$  &  $-0.0038$ & $-0.018$  \\
    &  $27$  &  $-0.00071$  & $-0.0034$ \\ \hline
    $PV$  &  $8_A$  &  $0.032$ & $0.15$\\
    &  $8_S$  &  $0.031$  & $0.15$ \\
    &  $10$  &  $0.020$ & $0.10$ \\
    &  $\overline{10}$  &  $0.016$ & $0.08$ \\
    &  $27$  &  $0.04$  & $0.19$\\ \hline
    $(VV)_{\mbox{$s$-wave}}$  &  $8$  &  $-0.081$ & $-0.39$ \\
    &  $27$  &  $-0.061$ & $-0.30$\\
    $(VV)_{\mbox{$p$-wave}}$  &  $8$  &  $-0.10$ & $ -0.48$\\
    &  $27$  &  $-0.14$ & $-0.70$ \\
    $(VV)_{\mbox{$d$-wave}}$  &  $8$  &  $0.51$ & $2.5$ \\
    &  $27$  &  $0.57$  & $2.8$\\ \hline
$(3P)_{\mbox{$s$-wave}}$        &  $8$  &  $-0.48$  & $-2.3$\\
    &  $27$  &  $-0.11$  & $-0.54$ \\
$(3P)_{\mbox{$p$-wave}}$        &  $8$  &  $-1.13$  & $-5.5$ \\
    &  $27$  &  $-0.07$   & $-0.36$  \\
$(3P)_{\mbox{form-factor}}$     &  $8$  &  $-0.44$  & $-2.1$\\
    &  $27$  &  $-0.13$ & $-0.64$ \\ \hline
$4P$  &  $8$  &  $3.3$ & $16$  \\
    &  $27$  &  $2.2$  & $11$ \\
    &  $27'$  &  $1.9$ & $9.2$ \\ \hline
\end{tabular} \vspace{4pt}
\caption{Values of $y_{F,R}$ for some two-, three-, and four-body final states.}
\label{ytwobody}
\end{center}
\end{table}

We observe that there are terms in Eq.~(\ref{ycombine}), like nonresonant $4P$, 
which could make contributions  to $y$ at the level of a percent or larger.  
There, the rest masses of the final state particles take up most of the available 
energy, so phase space differences are very important. One can see
that $y$  on the order  of a few percent is completely natural, and that anything 
an order of magnitude smaller would require significant  cancellations which do not
appear naturally in this framework.  Cancellations would be expected only if
they were enforced by the OPE, or if the charm quark were heavy
enough that the ``inclusive'' approach were applicable. The hypothesis
underlying the present analysis is that this is not the case.

\section{Conclusions}

We proved that if $SU(3)$ violation may be treated perturbatively,
then $D^0 - \D0bar$ mixing in the Standard Model is generated only at
second order in $SU(3)$ breaking effects.
Within the exclusive approach, we identified an $SU(3)$ breaking effect,
$SU(3)$ violation in final state phase space, which can be calculated
with minimal model dependence.  We found that phase space effects
alone provide enough $SU(3)$ violation to induce $y\sim10^{-2}$.
Large effects in $y$ appear for decays close to $D$ threshold, where
an analytic expansion in $SU(3)$ violation is no longer possible.

Indeed, some degree of cancellation is possible between different
multiplets, as would be expected in the $m_c\to \infty$ limit, or between
$SU(3)$ breaking in phase space and in matrix elements.  It is not known
how effective these cancellations are, and the most reasonable assumption 
in light of our analysis is that they are not significant enough to result 
in an order of magnitude suppression of $y$, as they are not enforced by any
symmetry arguments. Therefore, any future discovery of a $D$ meson width 
difference should not by itself be interpreted as an indication of the 
breakdown of the Standard Model.

It is my pleasure to thank S. Bergmann, E. Golowich, Y. Grossman, A. Falk, Z. Ligeti, 
and Y. Nir for collaboration on this and related projects.


\end{document}